\documentclass[preprint,showpacs,preprintnumbers,amsmath,amssymb]{revtex4}

\usepackage{graphicx}% Include figure files
\usepackage{dcolumn}% Align table columns on decimal point
\usepackage{bm}% bold math

\begin{document}

%\preprint{APS/123-QED}

\title{Experimental study on Gaussian-modulated coherent states quantum key distribution over standard telecom fiber}% Force line breaks with \\

\author{Bing Qi, Lei-Lei Huang, Li Qian, Hoi-Kwong Lo}

\affiliation{ Center for Quantum Information and Quantum Control
(CQIQC),
Dept. of Physics and Dept. of Electrical and Computer Engineering,\\
University of Toronto, Toronto, M5S 3G4, Canada }


\begin{abstract}
In this paper, we present a fully fiber-based one-way Quantum Key
Distribution (QKD) system implementing the Gaussian-Modulated
Coherent States (GMCS) protocol. The system employs a double
Mach-Zehnder Interferometer (MZI) configuration in which the weak
quantum signal and the strong Local Oscillator (LO) go through the
same fiber between Alice and Bob, and are separated into two paths
inside Bob's terminal. To suppress the LO leakage into the signal
path, which is an important contribution to the excess noise, we
implemented a novel scheme combining polarization and frequency
multiplexing, achieving an extinction ratio of 70dB. To further
minimize the system excess noise due to phase drift of the double
MZI, we propose that, instead of employing phase feedback control,
one simply let Alice remap her data by performing a rotation
operation. We further present noise analysis both theoretically and
experimentally. Our calculation shows that the combined polarization
and frequency multiplexing scheme can achieve better stability in
practice than the time-multiplexing scheme, because it allows one to
use matched fiber lengths for the signal and the LO paths on both
sides of the double MZI, greatly reducing the phase instability
caused by unmatched fiber lengths. Our experimental noise analysis
quantifies the three main contributions to the excess noise, which
will be instructive to future studies of the GMCS QKD systems.
Finally, we demonstrate, under the ``realistic model'' in which Eve
cannot control the system within Bob's terminal, a secure key rate
of 0.3bit/pulse over a 5km fiber link. This key rate is about two
orders of magnitude higher than that of a practical BB84 QKD system.

\end{abstract}

\maketitle

\section{Introduction}

One important practical application of quantum information is
quantum key distribution (QKD), whose unconditional security is
based on the fundamental laws of quantum mechanics
\cite{BB84,A91,Gisin02,securityproof}. In principle, any
eavesdropping attempts by a third party, Eve, will unavoidably
introduce quantum disturbances and be caught by the legitimate users
Alice and Bob.

Recently Gaussian-modulated coherent states (GMCS) QKD protocol has
drawn a lot of attention because of its potential high key rates,
especially over short distances
\cite{GMCS_NATURE,GMCS_PRA,GMCS_QIC,GMCS_NEW,GMCS_SECURITY}.
Compared with single photon QKD protocol (such as the BB84 QKD
\cite{BB84}), GMCS QKD protocol has several distinctive advantages:
First, the coherent state required in the GMCS QKD protocol can be
easily produced by a practical laser source; whereas, a single
photon source prescribed by the BB84 QKD is still unavailable. To
use a weak coherent source in a single photon QKD system, special
techniques, such as decoy states
\cite{DECOY_THEORY1,DECOY_THEORY2,DECOY_EXP}, are required to
improve the secure key rate. Second, the homodyne detectors in the
GMCS QKD protocol can be constructed using highly efficient PIN
diodes, while the performance of the single photon QKD is limited by
the low efficiency of today's single photon detector \cite{SSPD}.
Third, in GMCS QKD, information is encoded on continuous variables.
More than one bit of information could be transmitted by one pulse
and thus yields a high key rate.

Recent interest has also been sparked by the fact that
\cite{GMCS_NATURE}, with a ``reverse reconciliation'' protocol, GMCS
QKD can tolerate high channel loss ($>3$dB) on the condition that
the excess noise (the noise above vacuum noise) is not too high
($<0.5$). We remark that the security analysis given by
\cite{GMCS_NATURE} is applicable to individual attacks only. The
security of GMCS QKD protocol under the most general attack is still
under investigation \cite{GMCS_SECURITY}.

Despite its many advantages, the implementation of the GMCS QKD over
a practical distance in fiber remains challenging, and only one
other experimental demonstration has been reported so far
\cite{GMCS_NEW}. The major experimental challenge lies in the
reduction of the excess noise in a practical system. Here we study
the performance of a fully fiber-based one-way GMCS QKD system over
a 5km span. The purpose of this study is not only to show that GMCS
QKD can be operated over a practical distance, but also to
investigate various sources of excess noise in a real system, and to
offer practical solutions to reduce or eliminate some of the noise
sources. Our experiment with a 5km fiber demonstrates a secure rate
of 0.3bit/pulse under a ``realistic model'' in which we assume that
Eve cannot control Bob's system. This key rate is about two orders
higher than that of a practical BB84 QKD system.

This paper is organized as follows: Section II is a brief review of
GMCS QKD protocol. In Section III, we discuss our experimental setup
and summarize the experimental results. In Section IV, we present a
detailed noise analysis and discuss noise control in a practical
system. Section V is a brief conclusion.

\section{Gaussian-modulated coherent states (GMCS) QKD protocol}

The basic scheme of the GMCS QKD protocol is as follows
\cite{GMCS_NATURE}: Alice draws two random numbers $X_{A}$ and
$P_{A}$ from a set of Gaussian random numbers (with a mean of zero
and a variance of $V_{A}N_{0}$) and sends a coherent state $|X_{A} +
iP_{A}\rangle$ to Bob. Here $N_{0}=1/4$ denotes the shot-noise
variance \cite{SHOTNOISE}. In this paper, all variances are in the
shot noise units. Bob randomly chooses to measure either the
amplitude quadrature (X) or phase quadrature (P) with a phase
modulator and a homodyne detector. After performing his measurement,
Bob informs Alice which quadrature he actually measures for each
pulse through an authenticated public channel. Alice drops the
irrelevant data and only keeps the quadrature that Bob has measured.
At this stage, Alice shares a set of correlated Gaussian variables
(called the ``raw key'') with Bob. Alice and Bob then publicly
compare a random sample of their raw key to evaluate the
transmission efficiency of the quantum channel and the excess noise
of the QKD system.  Based on the above parameters, they can evaluate
the mutual information $I_{AB}$ and $I_{BE}$.

Assuming Alice's modulation variance is $V_{A}$, the channel
efficiency is $G$ and the total efficiency of Bob's device
(including the optical losses and the efficiency of homodyne
detector) is $\eta$, $I_{AB}$ and $I_{BE}$ are determined by
\cite{GMCS_NATURE}
\begin{align}
%\begin{equation}
I_{AB} =\frac{1}{2}\log_2 [(V+\chi)/(1+\chi)]\\
%\end{equation}
%\begin{equation}
I_{BE} =\frac{1}{2}\log_2[(\eta G)^{2}(V+\chi)(V^{-1}+\chi)]
%\end{equation}
\end{align}
Here, $V=V_{A}+1$ is the quadrature variance of the coherent state
prepared by Alice. $\chi$ is the equivalent noise measured at the
input, which can be separated into ``vacuum noise''
$\chi_{vac}=(1-\eta G)/\eta G$ (noise associated with the channel
loss and detection efficiency of Bob's system) and ``excess noise''
$\varepsilon$ (noise due to the imperfections in a non-ideal QKD
system):
\begin{align}
%\begin{equation}
\chi=\frac{1-\eta G}{\eta G}+\varepsilon
%\end{equation}
\end{align}

Assuming a reverse reconciliation algorithm efficiency of $\beta$,
the secure key rate is then given by \cite{GMCS_NATURE}
\begin{align}
%\begin{equation}
\Delta I=\beta I_{AB}-I_{BE}
%\end{equation}
\end{align}

Note, in (2), we assume that losses and noise in Bob's system can be
controlled by the eavesdropper Eve. In practice, it may be
reasonable to assume that Eve cannot control devices inside Bob's
system. Under this ``realistic model'' \cite{GMCS_NATURE}, noise
inside and outside of Bob's system are treated differently: while
part of the excess noise (e.g., due to imperfections outside of
Bob's system) might originate from  Eve's attack, the noise
contributed by Bob's devices is an intrinsic parameter of the QKD
system of which Eve has no control. Thus it is useful to write the
total excess noise $\varepsilon$ as
\begin{align}
%\begin{equation}
\varepsilon=\varepsilon_{A}+\frac{N_{Bob}}{\eta G}
%\end{equation}
\end{align}
where $\varepsilon_{A}$ denotes noise contribution from outside of
Bob's system, and $N_{Bob}$ denotes noise generated within Bob's
system (measured at the output). $\varepsilon_{A}$ and $N_{Bob}$ can
be determined separately.

From (3) and (5), the equivalent input noise is
\begin{align}
%\begin{equation}
\chi=\frac{1-\eta G}{\eta G}+\varepsilon_{A}+\frac{N_{Bob}}{\eta G}
%\end{equation}
\end{align}
Bob's quadrature variance is given by $V_{B}=\eta G(V+\chi)$, while
the conditional variance under the ``realistic model'' is
\begin{align}
%\begin{equation}
V_{B|E}=\frac{\eta}{1-G+G(\varepsilon_{A}+V^{-1})}+(1-\eta)+N_{Bob}
%\end{equation}
\end{align}
From (5)-(7), the mutual information $I_{BE}$ is
\begin{align}
%\begin{equation}
I_{BE} =\frac{1}{2}\log_2[\frac{\eta GV_{A}+1+\eta
G\varepsilon}{\eta/(1-G+G\varepsilon_{A}+GV^{-1})+1-\eta+N_{Bob}}]
%\end{equation}
\end{align}
Again, the secure key rate is determined by (4). Note (8) is
equivalent to (3) in \cite{GMCS_NEW}.

\section{GMCS-QKD experimental setup and experimental results}

In this section, we first present our experimental setup, followed
by discussions on the technical challenges. Finally, we present our
QKD experimental results.

\subsection{Experimental setup}
The schematic of our experimental setup is shown in Fig.1. The laser
source is a 1550nm continuous-wave fiber laser (NP Photonics). Alice
uses a LiNbO$_{3}$ amplitude modulator (AM$_{0}$) to generate 200-ns
laser pulses at a repetition rate of 100KHz. She then prepares a
coherent state $|X_{A} + iP_{A}\rangle$ with the second amplitude
modulator (AM$_{1}$) and a phase modulator (PM$_{1}$). AM$_{1}$ and
PM$_{1}$ are driven by Arbitrary Waveform Generators (AWG) which
contain random amplitude and phase data produced from $\{X_{A},
P_{A}\}$. Alice sends Bob the quantum signal together with a strong
local oscillator (LO) as the phase reference through a 5km telecom
fiber.  On Bob's side, he randomly chooses to measure either $X$ or
$P$ with his phase modulator (PM$_{2}$) and a homodyne detector. The
phase modulator PM$_{2}$ is located in the reference path of Bob's
MZI and is driven by a third AWG which contains a binary random file
for choosing $X$ or $P$. The homodyne detector is constructed by a
pair of photo-diodes and a low noise charge sensitive amplifier,
similar to the one described in \cite{HOMODYNE}. Note, to reduce the
noise due to multiple reflections of LO in Bob's system, a fiber
isolator has been placed in the signal arm of Bob's MZI. The outputs
of the homodyne detector are sampled by a 12-bit data acquisition
card (NI, PCI-6115) at a sampling rate of 10MS/s.

\begin{figure}[!t]\center
\resizebox{12.0cm}{!}{\includegraphics{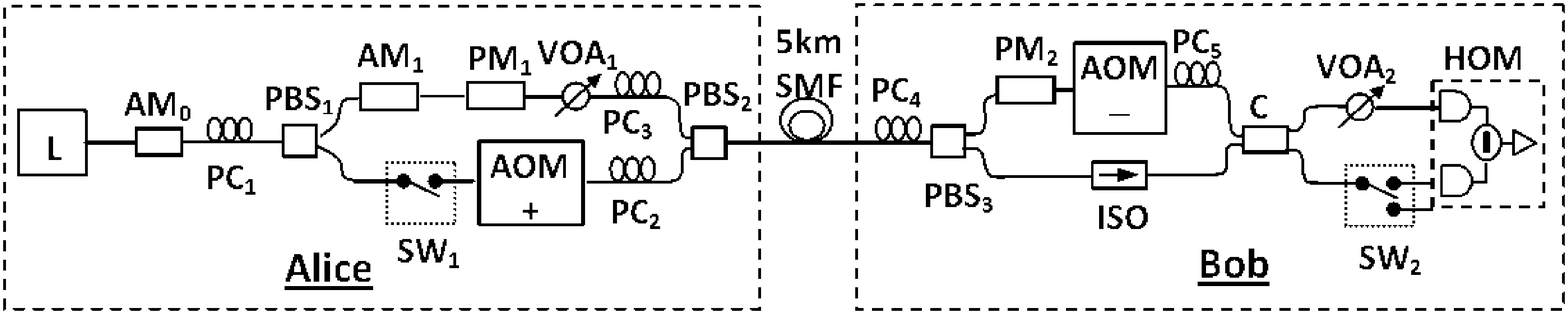}} \caption{The
optical layout of our GMCS QKD system. L: 1550 nm CW fiber laser;
PC$_{1-5}$: polarization controllers; PBS$_{1-3}$: polarization beam
splitters/combiners; AM$_{0-1}$: amplitude modulators; PM$_{1-2}$:
phase modulators; SW$_{1-2}$: optical switches; AOM$+$ (AOM$-$):
upshift (downshift) acousto-optic modulator; VOA$_{1-2}$: variable
optical attenuators; ISO: isolator; C: fiber coupler; HOM: homodyne
detector.}
\end{figure}

There are two significant technical challenges in this double
Mach-Zehnder interferometer (MZI) scheme: First, the leakage (LE) of
the strong LO (typically $10^{8}$ photons/pulse) into the signal
path has to be reduced effectively, particularly because the quantum
signal is very week (typically less than 100photons/pulse). Ideally
there should be no LE. The LO and the signal (Sig) are supposed to
go through different arms in Bob's interferometer. For an non-ideal
system in our experiment, however, we expect that there will be some
leakage LE to the same arm as the signal, see Fig.2. If LE is in the
same spatiotemporal mode and the same polarization state as the LO,
it will interfere with LO and contribute to the excess noise.
Second, the phase fluctuation introduced by the MZI, which is one of
the major contributions to excess noise, has to be minimized. We
discuss these issues in the next two subsections.

\begin{figure}[!t]\center
\resizebox{10cm}{!}{\includegraphics{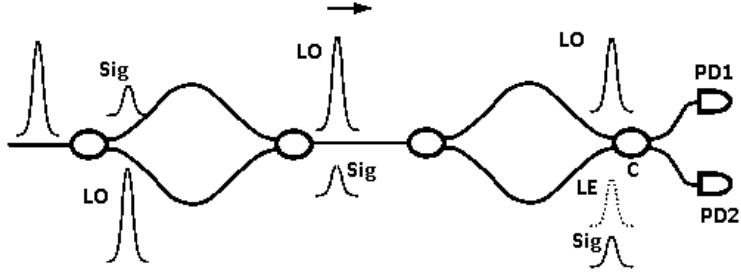}} \caption{The
leakage of the local oscillator in the double Mach-Zehnder
interferometer scheme: Sig-quantum signal; LO-local oscillator;
LE-leakage of LO.}
\end{figure}

\subsection{Reduce the leakage of the local oscillator}

In a report by J. Lodewyck, et al. \cite{GMCS_NEW}, to reduce excess
noise due to the leakage, LE is separated from LO in the time domain
by using MZIs with largely unbalanced path lengths. Since LE and LO
arrive at the fiber coupler ($C$ in Fig.1) at different times, they
interfere with each other only weakly. Obviously, to minimize the
overlap between LE and LO in the time domain, the required time
delay should be much larger than the width of the laser pulse. This
corresponds to a large length unbalance in the MZI (In
\cite{GMCS_NEW}, the length unbalance of MZI is 80m). However, it is
quite challenging to stabilize a MZI with such a large length
unbalance in a practical system. The phase fluctuation of the
unbalanced MZI may result in a dramatic increase in the excess
noise.

In contrast, we employ polarization multiplexing combined with
frequency multiplexing to minimize the leakage of the LO. Alice uses
orthogonal polarization states for the quantum signal and the LO via
a polarization beam splitter (PBS$_{1}$ in Fig.1). On Bob's side,
another polarization beam splitter (PBS$_{3}$ in Fig.1) is used to
separate the LO from the signal. This polarization multiplexing
scheme is expected to yield an extinction ratio of about 30dB due to
the imperfections of the PBSs. To further suppress the excess noise
due to the leakage, we have introduced a frequency multiplexing
technique: a pair of acousto-optic modulators (AOM$+$ and AOM$-$ in
Fig.1) are used to upshift and downshift the frequency of the LO by
55MHz. As a result, the majority of LE can be filtered out since it
has a different frequency from LO. Although in principle the phase
of the LO will also be shifted by the AOM \cite{AOM_OE}, since the
driving frequency of the AOM (55MHz) is much smaller than the laser
frequency (200THz), the phase noise contributed by the AOM is
negligible.

The overall equivalent extinction ratio of this scheme has been
determined experimentally to be around 70dB, and the excess noise
due to the leakage is about 0.02 (measured at the output, see
details in Section IV).

\subsection{Reduce phase fluctuation of the MZI}

In both the GMCS QKD system and the phase coding BB84 QKD system,
ideally, the phase difference between the quantum signal and the LO
(phase reference) should be solely dependent on the phase
information encoded by Alice. However, in practice, the zero point
of the phase difference $\phi_{0}$ (the phase difference when Alice
encodes phase 0) will drift with time. The GMCS QKD protocol is more
sensitive to this phase drift than the BB84 QKD protocol in the
sense that a small phase drift would lower the secure key rate
dramatically \cite{GMCS_PHASE}.

Under normal condition, $\phi_{0}$ drifts with time slowly. It is
reasonable to assume that $\phi_{0}$ is constant during one frame of
QKD transmission (40ms in our experiment). As shown in Fig.3, the
change of $\phi_{0}$ measured during the QKD is $0.016/s$, or
$6.4\times10^{-4}$ in 40ms. The corresponding contribution to excess
noise (with a modulation variance of $16.9$) is about
$7\times10^{-6}$, which is negligible. Alice and Bob can estimate
the value of $\phi_{0}$ in this transmission period by comparing a
subset of their QKD data.

\begin{figure}[!t]\center
\resizebox{10cm}{!}{\includegraphics{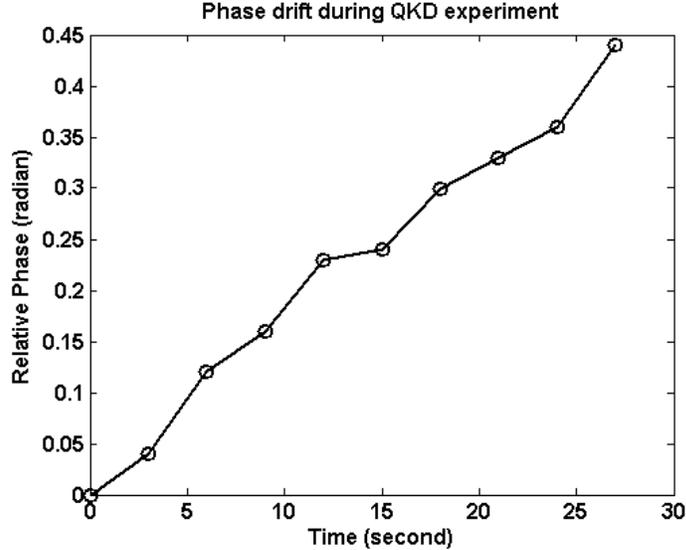}} \caption{The phase
drift observed during QKD experiment without active phase
stabilization. Each point in the curve is estimated from the QKD
data in 40ms (Here, we assume the phase change in 40ms is small
enough to be neglected). The total phase drift is about $0.016/s$,
or $6.4\times10^{-4}$ in 40ms.}.
\end{figure}

In phase coding BB84 QKD system, knowing the value of $\phi_{0}$
itself will not help Alice and Bob to lower the quantum bit error
rate (QBER). To control the QBER due to the phase drift, a phase
re-calibration process is essential: Alice and Bob have to perform a
phase feedback control to compensate this phase drift before they
start the key transmission \cite{PHASE_CONTROL}.

In contrast, in GMCS QKD, we propose a simpler way to remove the
excess noise due to the phase drift $\phi_{0}$: once Alice and Bob
know the value of $\phi_{0}$, instead of performing feedback phase
control, Alice can simply modifiy her data to incorporate this phase
drift. Specifically, during the classical communication stage, Bob
announces a randomly-selected subset of his measurement results.
Alice can estimate $\phi_{0}$ and other system parameters from Bob's
measurement results and her original data. Then she maps her data
$\{X_{A}, P_{A}\}$ into $\{X'_{A}, P'_{A}\}$ by performing
\begin{align}
%\begin{equation}
X'_{A}=X_{A}\cos{\phi_{0}}+P_{A}\sin{\phi_{0}}\\
P'_{A}=-X_{A}\sin{\phi_{0}}+P_{A}\cos{\phi_{0}}
%\end{equation}
\end{align}
Alice and Bob can produce a secure key from $\{X'_{A}, P'_{A}\}$ and
$\{X_{B}, P_{B}\}$. The security analysis of GMCS QKD still holds.

The above approach reduces the excess noise due to the slow drift of
$\phi_{0}$, but it does not solve the problem of fast variations in
$\phi_{0}$ resulted from instabilities in the MZIs. This instability
is worse when the path lengths of the MZIs are not balanced.
Fortunately, because we employ the combined polarization and
frequency multiplexing instead of time multiplexing, we can use
balanced MZIs. To further stabilize the MZIs, we carefully balance
their path lengths and place each of them into an enclosure to
minimize environmental noise.

\subsection{Experimental results}

We perform the QKD experiment with a strong LO ($8\times10^{7}$
photons/pulse) and a signal of modulation variance of $16.9$. Data
are transmitted by frames. Each frame contains $4000$ points
(Gaussian random numbers). Among them, Bob performs $X$ quadrature
measurements on $1980$ points and $P$ quadrature measurements on
$2020$ points. The same random patterns are used repeatedly in our
experiment. The experimental results are shown in Fig.4a. The
equivalent input noise has been determined experimentally to be
$\chi=2.25$. For comparison, Fig.4b shows the simulation results
under the assumption of no excess noise.
\begin{figure}[!t]\center
\resizebox{10cm}{!}{\includegraphics{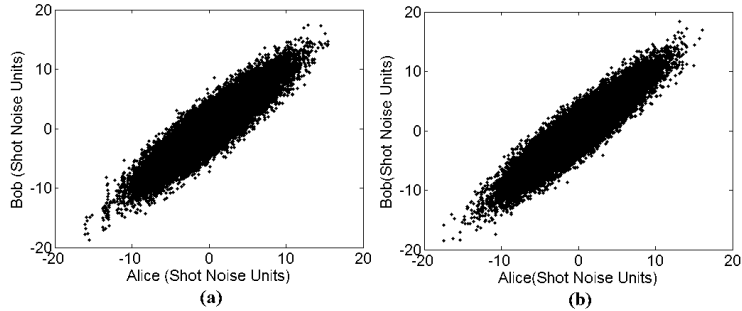}} \caption{(a) QKD
experimental results ($40000$ points). The equivalent input noise
has been determined experimentally to be $\chi=2.25$, which includes
``vacuum noise'' $\chi_{vac}=2.00$ and ``excess noise''
$\varepsilon=0.25$. (b) Simulation results: assuming ``vacuum
noise'' $\chi_{vac}=2.00$ and ``excess noise'' $\varepsilon=0$. }.
\end{figure}

The channel efficiency $G$ and the total efficiency of Bob's device
$\eta$ have been calibrated carefully to be $G=0.758$ and $\eta
=0.44$ (including optical loss in Bob's system $0.61$ and the
efficiency of the homodyne detector $0.72$)\cite{CALIBRATION}. Using
(1), (2) and (4) the secure key rate under the general model has
been calculated to be 0 if we assume $\beta=0.898$ \cite{GMCS_NEW}
or 0.13bit/pulse is we assume $\beta=1$.

To estimate the secure key rate under the ``realistic model'', we
need to determine $\varepsilon$, $\varepsilon_{A}$ and $N_{Bob}$.
From $\chi=2.25$, $G=0.758$ and $\eta =0.44$, we can determine
$\varepsilon=0.25$ by using (3). Experimentally, as will be
discussed in detail in Section IVA later, we have estimated
$\varepsilon_{A}=0.056$. From (5), we can calculate $N_{Bob}=0.065$
(see details in Section IV). Using (1), (8) and (4), the secure key
rate under the realistic model has been calculated to be either
$0.30$ ($\beta=0.898$) or 0.43 ($\beta=1$). Table 1 summarizes our
experimental results.

\begin{table}[!b]\center
\caption{QKD parameters and results (e: experimental result; c:
calculated result).}\label{Tab: Experimental Result}
\begin{tabular}{c c c c c c c c c c c c}
\hline $V_{A}$ & $G$ & $\eta$ & $\chi$ & $\varepsilon$ & $\varepsilon_{A}$ & $N_{Bob}$ & $R^{gen}_{\beta=1}$ & $R^{gen}_{\beta=0.898}$ & $R^{rea}_{\beta=1}$ & $R^{rea}_{\beta=0.898}$\\
\hline 16.9(e) & 0.758(e) & 0.44(e) & 2.25(e) & 0.25(c) & 0.056(e) & 0.065(c) & 0.13(c) & 0(c) & 0.43(c) & 0.30(c)\\
\hline
\end{tabular}
\end{table}

Using the parameters in Table 1, we have performed numerical
simulations under both the general model and the realistic model.
Here we assume the quantum channel is telecom fiber with a loss of
$0.21$ dB/km. Fig.5a shows the result with a perfect reverse
reconciliation algorithm ($\beta=1$). Fig.5b shows the result with a
practical reverse reconciliation algorithm ($\beta=0.898$).
\begin{figure}[!t]\center
\resizebox{10cm}{!}{\includegraphics{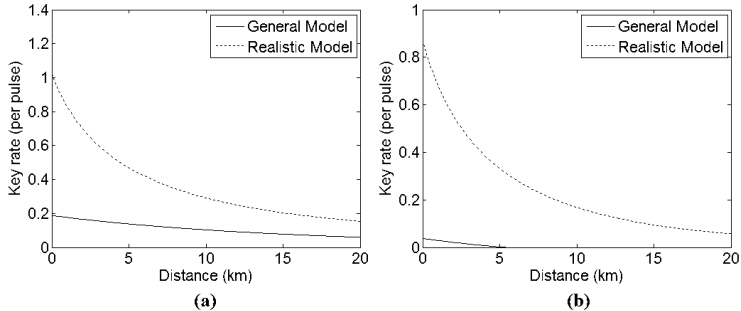}}
\caption{Simulation result results (a)$\beta=1$ (b)
$\beta=0.898$\cite{GMCS_NEW}}.
\end{figure}

As shown in Fig.5, under the ``realistic model'', the achievable
secure key rate is significantly higher than that of a practical
BB84 QKD.

\section{Experimental investigation and analysis on excess noise}

To estimate the secure key rate under the ``realistic model'', we
have to separate $\varepsilon$ into $\varepsilon_{A}$ and $N_{Bob}$
(see (8)). In this section, we will discuss how to estimate
$\varepsilon_{A}$ and $N_{Bob}$ in a practical GMCS QKD system and
other practical issues.

\subsection{Estimate $\varepsilon_{A}$}

$\varepsilon_{A}$ is the excess noise due to imperfections outside
of Bob's system, which includes the phase noise of the laser source,
imperfect amplitude and phase modulations, the phase noise of the
interferometer, etc.. To reduce the phase noise of MZIs, we
carefully balance their path lengths and enclose them to minimize
environmental noise. To reduce the excess noise due to the imperfect
modulations, both the amplitude modulator and the phase modulator
have been calibrated carefully before the QKD experiment.
Nevertheless, Alice and Bob have to measure $\varepsilon_{A}$
experimentally in order to apply the``realistic model''.

Following \cite{GMCS_PRA}, we assume that $\varepsilon_{A}$ is
proportional to the modulation variance $V_{A}$ and can be described
by $\varepsilon_{A}=V_{A}\delta$ . We have designed a procedure to
determine the proportionality constant $\delta$, by operating the
system with a large modulation variance ($V_{A}\approx40000$) and a
weak LO ($10^{5}$ photons/pulse, to reduce its leakage). Under this
condition, all other excess noises in (6) except $\varepsilon_{A}$
are negligible, ie., $\chi\simeq V_{A}\delta$. We can determine
$\delta$ by normalizing the observed equivalent input noise $\chi$
to the modulation variance $V_{A}$.

Fig.6 shows the experimental results. The measured $\delta$ is
$0.0033$ (In another test with $V_{A}\approx80000$, the measured
$\delta$ is 0.0032). Therefore, for a modulation variance of
$V_{A}=16.9$, the expected excess noise component
$\varepsilon_{A}=0.056$.

\begin{figure}[!t]\center
\resizebox{10cm}{!}{\includegraphics{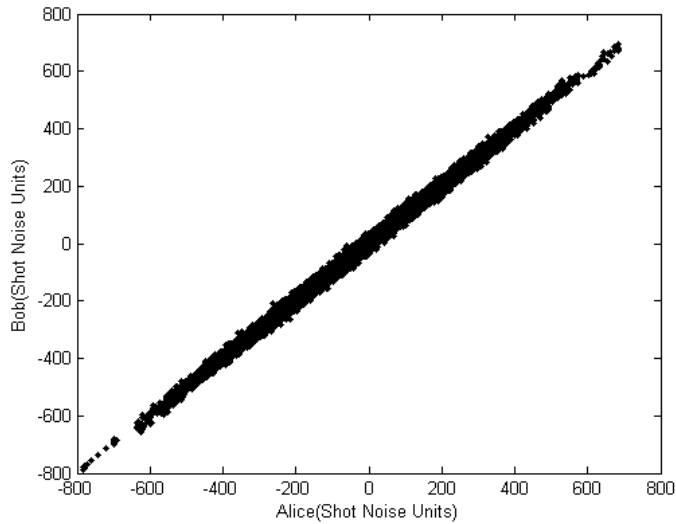}} \caption{Determine
$\delta$ by using a high modulation variance $V_{A}\approx40000$ and
a weak LO ($10^{5}$ photons per pulse). The result is
$\delta=0.0033$ ($40000$ points).}
\end{figure}

\subsection{Estimate $N_{Bob}$}

In Section IIID and IIIA, we experimentally determined: $\chi=2.25$,
$G=0.758$, $\eta =0.44$, and $\varepsilon_{A}=0.056$. From these
parameters, we can obtain $\varepsilon=0.25$ by using (3), and
obtain $N_{Bob}=0.065$ by using (5).

In this subsection, we will discuss the two main sources of
$N_{Bob}$, namely, the electrical noise of the homodyne detector
($N_{el}$) and the noise associated with the leakage of LO
($N_{leak}$).

Since the electrical noise of the homodyne detector scales with its
bandwidth, intuitively, a narrow bandwidth should be used to
minimize the electrical noise. However, a narrow bandwidth would
result in a wide pulse in time domain, which in turn reduces the
achievable repetition rate of the QKD system. Therefore, a trade-off
has to be made between the speed and the electrical noise.

We remark that this constraint on the noise and the speed of the
homodyne detector could be relaxed by adopting the ``dual-detector
method'' \cite{DUALDETECTOR}: the legitimate receiver randomly uses
either a fast but noisy detector or a quiet but slow detector to
measure the incoming quantum signals. The measurement results from
the quiet detector can be used to upper bound the eavesdropper's
information, while the measurement results from the fast detector
are used to generate a secure key.

Nevertheless, in our current setup, the bandwidth of the homodyne
detector is about 1MHz. The electrical noise is about 13.4dB below
the shot noise (with a LO of $8\times10^{7}$ photons/pulse), as
shown in Fig.7. The corresponding $N_{el}$ is therefore $0.045$.

\begin{figure}[!t]\center
\resizebox{10cm}{!}{\includegraphics{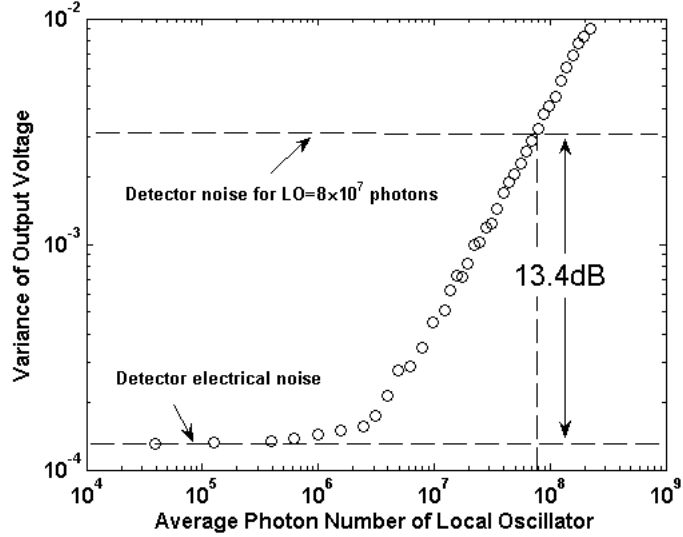}} \caption{Noise of
the balanced homodyne detector. The electrical noise is independent
of the photon number of the local oscillator while the shot noise is
directly proportional to the photon number of the local oscillator.
With a local oscillator of $8\times10^{7}$ photons/pulse, the
electrical noise (the variance observed at a low photon number of
the local oscillator) is about 13.4dB below the shot noise .}
\end{figure}

The analysis of the excess noise associated with the leakage of LO
is more complicated. Here, we estimate the order of magnitude of
$N_{leak}$ in both time-multiplexing scheme and
polarization-frequency-multiplexing scheme by treating the leakage
LE as a classical electromagnetic wave with a Gaussian shape
\cite{SEMICLASSICAL}. More rigorous results could be acquired by
solving this problem quantum mechanically.

\textbf{Case 1: $N_{leak}$ in time-multiplexing scheme}

In this scheme, MZIs with large unbalanced paths are employed to
introduce a time delay between the LO and its leakage LE, as shown
in Fig.8. We denote the average photon number of the leakage as
$\langle n_{le}\rangle$. Note only part of LE--the part that is in
the same spatiotemporal mode as the LO--will interfere with LO and
contribute to the excess noise. We denote the average photon number
of this ``effective'' leakage as $\langle n^{e}_{le}\rangle$.

\begin{figure}[!t]\center
\resizebox{10cm}{!}{\includegraphics{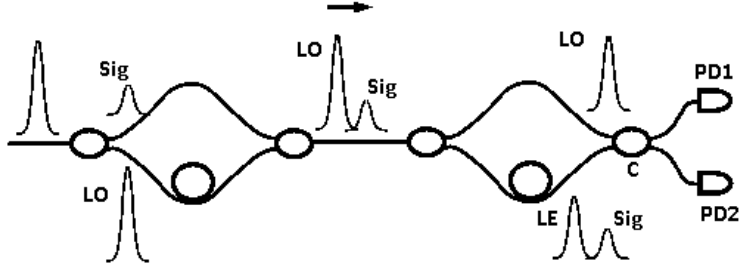}} \caption{The
time-multiplexing scheme: Sig-quantum signal; LO-local oscillator;
LE-leakage of LO. Note Sig and LO arrive the fiber coupler (C) at
the same time, while LE has been delayed.}
\end{figure}

The ``effective'' leakage $\langle n^{e}_{le}\rangle$ can be
estimated from
\begin{align}
%\begin{equation}
\langle n^{e}_{le}\rangle=\alpha\langle n_{le}\rangle
%\end{equation}
\end{align}
where $\alpha$ is the overlapping factor between LO and LE.

Assuming a Gaussian pulse shape, the normalized electrical fields of
LO and LE can be described by
\begin{align}
%\begin{equation}
E_{lo}=E_{0}\exp(-\frac{(t-\Delta_{t}/2)^{2}}{2\sigma^{2}_{t}})\exp(-i\omega_{0}t)\\
E_{le}=E_{0}\exp(-\frac{(t+\Delta_{t}/2)^{2}}{2\sigma^{2}_{t}})\exp[-i(\omega_{0}t+\phi_{le})]
%\end{equation}
\end{align}
Here the normalizing factor is
$E_{0}^{2}=\frac{1}{\sqrt{\pi}\sigma_{t}}$, $\Delta_{t}$ is the time
delay between LO and LE, $\phi_{le}$ is the phase difference between
LO and LE, and $\sigma_{t}$ is related to the full width at half
maximum (FWHM) $\sigma_{FW}$ by
$\sigma_{t}=\frac{\sigma_{FW}}{2\sqrt{ln2}}$.

The overlapping factor $\alpha$ can be calculated from
\begin{align}
%\begin{equation}
\alpha=|\int_{-\infty}^{\infty}E_{lo}^{\ast}E_{le}dt|^{2}=[E_{0}^{2}\int_{-\infty}^{\infty}\exp(-\frac{t^{2}}{\sigma^{2}_{t}})dt]^{2}\exp(-\frac{\Delta_{t}^{2}}{2\sigma^{2}_{t}})=\exp(-\frac{\Delta_{t}^{2}}{2\sigma^{2}_{t}})
%\end{equation}
\end{align}
Here we use the normalization relation
$E_{0}^{2}\int_{-\infty}^{\infty}\exp(-\frac{t^{2}}{\sigma^{2}_{t}})dt=1$.

If Bob chooses to measure the X quadrature, the contribution from
the leakage is (see Fig.9)
\begin{align}
%\begin{equation}
X_{le}=\sqrt{\langle n^{e}_{le}\rangle}\cos\phi_{le}
%\end{equation}
\end{align}

\begin{figure}[!t]\center
\resizebox{10cm}{!}{\includegraphics{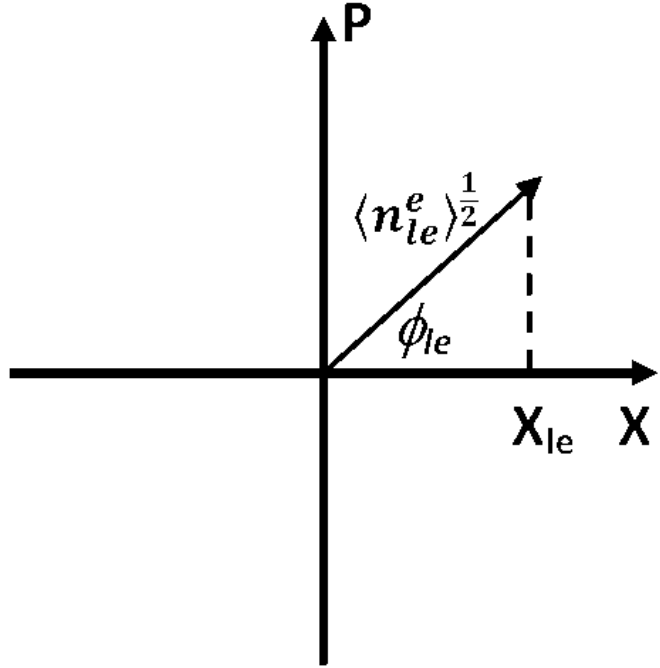}} \caption{The
contribution of the effective leakage on X quadrature measurement. }
\end{figure}

Because of the large length unbalance required in this scheme, we
assume that the relative phase $\phi_{le}$ randomly and rapidly
changes in the range of $[0,2\pi]$. The corresponding excess noise
(in shot noise units) is
\begin{align}
%\begin{equation}
N_{leak}=4\langle X_{le}^{2}\rangle=4\langle n^{e}_{le}\rangle
\langle(\cos\phi_{le})^{2}\rangle=2\langle n^{e}_{le}\rangle
%\end{equation}
\end{align}

Using (11), (14) and (16), $N_{leak}$ can be estimated by
\begin{align}
%\begin{equation}
N_{leak}=2\langle
n_{le}\rangle\exp(-\frac{\Delta_{t}^{2}}{2\sigma^{2}_{t}})
%\end{equation}
\end{align}

From another point of view, the required time delay for a given
$N_{leak}$ can be estimated by
\begin{align}
%\begin{equation}
\Delta_{t}=\sqrt{2\ln(\frac{2\langle
n_{le}\rangle}{N_{leak}})}\sigma_{t}
%\end{equation}
\end{align}

If a simple time-multiplexing scheme is adopted, and a 3dB coupler
is used in Bob's MZI, the leakage LE will be on the same order as
LO. Assuming $\langle n_{le}\rangle=10^{8}$, $\sigma_{t}=60$ns
(corresponds to $\sigma_{FW}=100$ns), to suppress the excess noise
$N_{leak}$ to below $0.02$ (this is the $N_{leak}$ observed in our
polarization-frequency-multiplexing setup), the required time delay
calculated from (18) is about 406ns, which corresponds to a 81m
fiber length difference in MZI.

If time-multiplexing and polarization-multiplexing are combined to
suppress the leakage, then the leakage LE will be three orders of
magnitude lower than LO (assuming a 30dB polarization extinction
ratio). Using $\langle n_{le}\rangle=10^{5}$, $\sigma_{t}=60$ns and
$N_{leak}=0.02$, the required time delay is about 340ns, which
corresponds to a 68m fiber length difference in MZI.

Based on the above calculations, we can see that although the excess
noise due to leakage can be effectively reduced by employing this
time multiplexing scheme, the required length unbalance is quite
large. In practice, it is quite challenging to stabilize a MZI with
such a large length unbalance. Without phase stabilization, the
phase fluctuation of the unbalanced MZI will result in a dramatic
increase in the excess noise.

\textbf{Case 2: $N_{leak}$ in polarization-frequency-multiplexing
scheme}

If the laser pulse has an ideal Gaussian-shaped spectrum, the
calculations in Case 1 can be easily extended into frequency domain.
Similar to (14), in the spectral domain, the overlapping factor
$\alpha$ can be estimated from
\begin{align}
%\begin{equation}
\alpha=\exp(-\frac{\Delta_{\nu}^{2}}{2\sigma^{2}_{\nu}})
%\end{equation}
\end{align}
where $\Delta_{\nu}$ is the frequency difference between LO and LE,
while $\sigma_{\nu}$ is the spectral width of the laser pulse.

For a 100ns (FWHM) transform limited Gaussian pulse, it's spectral
width (FWHM) is about 4.4MHz, or $\sigma_{\nu}\approx2.64$MHz. With
a $\Delta_{\nu}$ of 55MHz, from (19), we would expect an extremely
small $\alpha$ ($<10^{-90}$), which means the leakage contribution
to the excess noise is negligible. Though in practice, the spectrum
of a practical laser source doesn't have an ideal Gaussian shape:
far from the peak wavelength, the spectral power density approaches
a constant noise floor. The overlapping factor $\alpha$ is mainly
determined by this noise floor.

Here, we estimate the order of magnitude of $\alpha$ from
experimental data directly. Since we design MZIs with carefully
balanced path lengths, in the period of one frame of transmission
(40ms), the phase difference between LO and LE has a constant
average value $\phi_{le}^{(0)}$ with a small fluctuation term
$\Delta\phi_{le}$
\begin{align}
%\begin{equation}
\phi_{le}=\phi_{le}^{(0)}+\Delta\phi_{le}
%\end{equation}
\end{align}

Consequently, the contributions of LE to Bob's measurement results
are (see Fig.9)
\begin{align}
%\begin{equation}
X_{le}=\sqrt{\langle
n^{e}_{le}\rangle}\cos\phi_{le}=X_{le}^{(0)}+\Delta X_{le}\\
P_{le}=\sqrt{\langle
n^{e}_{le}\rangle}\sin\phi_{le}=P_{le}^{(0)}+\Delta P_{le}
%\end{equation}
\end{align}
where
\begin{align}
%\begin{equation}
X_{le}^{(0)}=\sqrt{\langle n^{e}_{le}\rangle}\cos\phi_{le}^{(0)}\\
P_{le}^{(0)}=\sqrt{\langle n^{e}_{le}\rangle}\sin\phi_{le}^{(0)}\\
\Delta X_{le}\approx-\sqrt{\langle n^{e}_{le}\rangle}(\sin\phi_{le}^{(0)})\Delta\phi_{le}\\
\Delta P_{le}\approx\sqrt{\langle
n^{e}_{le}\rangle}(\cos\phi_{le}^{(0)})\Delta\phi_{le}
%\end{equation}
\end{align}

Since $X_{le}^{(0)}$ and $P_{le}^{(0)}$ are constant in each frame,
Bob can determine their values from his experimental results and
remove their contributions by simply shifting his data. So
$X_{le}^{(0)}$ and $P_{le}^{(0)}$ will not contribute to excess
noise. In our QKD experiment, during the post-processing stage, Bob
calculates the DC component of his measurement results for each
transmission frame, then simply subtracts this DC component from his
original data.

In addition, the ``effective'' leakage $\langle n^{e}_{le}\rangle$
and $\phi_{le}^{(0)}$ can be estimated from experimentally obtained
$X_{le}^{(0)}$ and $P_{le}^{(0)}$:
\begin{align}
%\begin{equation}
\langle n^{e}_{le}\rangle=(X_{le}^{(0)})^2+(P_{le}^{(0)})^2\\
\phi_{le}^{(0)}=\arctan(\frac{P_{le}^{(0)}}{X_{le}^{(0)}})
%\end{equation}
\end{align}

During the QKD experiment, the average photon number of LO is around
$8\times10^{7}$, while the $\langle n^{e}_{le}\rangle$ has been
determined using (27) to be $6$, indicating an overall equivalent
extinction ratio of $\sim70$dB.

From (25) and (26), the excess noise due to leakage $N_{leak}$ is
proportional to $\langle n^{e}_{le}\rangle$ can be described by
\begin{align}
%\begin{equation}
N_{leak}=\langle n^{e}_{le}\rangle\gamma
%\end{equation}
\end{align}

Let us estimate $\gamma$ from experimental data: $N_{Bob}$ has been
determined to be $0.065$ (Section IVB) and $N_{el}$ has been
determined to be $0.045$ (Section IVB). Thus $N_{leak}$ is about
$0.02$. Using $\langle n^{e}_{le}\rangle\approx6$, we obtain
$\gamma$ to be on the order of $0.003$. In Section IVA, we described
$\varepsilon_{A}$ as $V_{A}\delta$ and determined $\delta$ to be
$0.0033$. Since both $\gamma$ and $\delta$ are associated with the
phase noise of MZI and the laser source, we expect that these
quantities to have the same order of magnitude, and indeed they do.

One major advantage of the polarization-frequency-multiplexing
scheme is that balanced MZIs can be employed. Under the same
conditions, the phase noise of balanced MZIs should be much lower
than MZIs with large path length imbalance. The resulting
improvements are two folds: first, a small phase fluctuation between
LO and signal corresponds to a small excess noise $\varepsilon_{A}$.
Secondly, a small phase fluctuation between LO and LE reduces the
excess noise due to the leakage.

\subsection{Other practical issues with GMCS QKD}

As shown in Table 1, under the ``realistic model'', the achievable
secure key rate of our system is significantly higher than that of a
practical BB84 QKD over short distances. However, to achieve such a
high key rate, the excess noises in the system need to be controlled
effectively and the system parameters need to be determined with
high accuracies.

Note in the BB84 QKD system with a single photon source, Eve's
information is upper bounded by the QBER, which can be estimated by
Alice and Bob from their QKD results directly. In practice, a
moderate error on determining QBER will not change the secure key
rate significantly \cite{BB84_KEY}.

However, there is a major challenge in GMCS QKD: to calculate the
secure key rate under the ``realistic model'', in addition to the
total transmission efficiency (which is the product of $G$, $\eta$
and the gain of Bob's electrical amplifier) and the equivalent input
noise $\chi$ (which can be determined from Bob's measurement
results), Alice and Bob have to develop techniques to monitor other
system parameters $V_{A}$, $G$, $\eta$ and $\varepsilon_{A}$ with
high degree of accuracy in real time.

For example, among the total equivalent input noise $\chi=2.25$, the
contribution of vacuum noise ($2.0$) is much higher than that of the
excess noise ($0.25$)\cite{FIG4}. To acquire a tight bound on
$\varepsilon_{A}$ from the experimentally measured equivalent input
noise $\chi$ (see (6)), Bob has to determine the total efficiency
$\eta G$ with an extremely high accuracy. Using (6) and parameters
in Table 1, to achieve an accuracy of $0.01$ in $\varepsilon_{A}$
estimation, the required accuracy on $\eta G$ estimation is $0.1\%$.

To estimate $\varepsilon_{A}$ accurately without referring to $\eta
G$, we have designed a separated calibration process (see Section
IVA). Strictly speaking, this cannot be applied to QKD experiment
directly, since Eve may attack this calibration process and QKD
process differently. We need to develop special techniques to
estimate each system parameter accurately without compromising the
security of the QKD system.

\section{Conclusion}

Gaussian-modulated coherent states (GMCS) quantum key distribution
(QKD) protocol has been proposed to achieve efficient secure key
distribution with standard telecommunication components. The
performance of a practical GMCS QKD system is mainly determined by
its excess noise. In this paper, we present a fully fiber GMCS-QKD
system based on double Mach-Zehnder interferometer (MZI) scheme and
build up a corresponding theoretical model for noise analysis. To
effectively reduce the excess noise due to the leakage from the
strong local oscillator to the weak quantum signal, we introduce a
novel polarization-frequency-multiplexing scheme. To minimize the
excess noise due to the phase drift of MZI, instead of using phase
feedback control, we propose that the sender simply remap her data
by performing a rotating operation. The experiment with a 5km fiber
demonstrates a secure key rate of 0.3bit/pulse under the ``realistic
model''. This secure key rate is about two orders higher than that
of a practical BB84 QKD system.

We analyzed and quantified various sources of excess noise in a
practical GMCS QKD system, and offered practical solutions to reduce
or eliminate some of the noise sources. We believe, in order to
achieve a high secure key rate in real world, special techniques for
estimating system parameters with high accuracies in real time
(without compromise the security of the QKD system) are in demand.
High speed GMCS QKD is also an important research direction for the
future.

We thank Ryan Bolen and Justin Chan for their work on the homodyne
detector and Alexander Lvovsky for helpful discussions. Financial
support from NSERC, CIFAR, CRC Program, CFI, OIT, MITACS, PREA, CIPI
and QuantumWorks are gratefully acknowledged. This research was
supported by Perimeter Institute for Theoretical Physics. Research
at Perimeter Institute is supported in part by the Government of
Canada through NSERC and by the province of Ontario through MEDT.


\begin{thebibliography}{00}
\bibitem{SSPD}
We remark that new types of single photon detector (SPD), such as
Superconductor SPD, could have high efficiency at telecom
wavelength.
\bibitem{GMCS_PHASE}
Assume that the phase drift of MZI $\phi_{0}<<1$. In the phase
coding BB84 QKD, the resulting QBER is roughly equal to
$\phi_{0}^{2}$. In practice, a $1\%$ error rate is acceptable (The
secure bound on QBER is about $20\%$ for two way classical
communication \cite{BB84_BOUND}) which corresponds to
$\phi_{0}=0.1$.

In GMCS QKD, Bob's measurement result of $X$ quadrature becomes
$X'=X\cos{\phi_{0}}+P\sin{\phi_{0}}$. If $\phi_{0}<<1$ and its
change during the time of one frame transmission ($40$ ms in our
experiment) is negligible, the excess noise contributed by the phase
drift can be estimated by $\langle(X'-X)^{2}\rangle\approx\langle
P^{2}\rangle\phi_{0}^{2}=V_{A}\phi_{0}^{2}$. With a modulation
variance of $V_{A}=20$, a $0.1$ phase drift will result an excess
noise of $0.2$. Note that the secure bound (of excess noise) for a
reverse reconciliation protocol is around $0.5$ \cite{GMCS_NATURE}.
An excess noise of $0.2$ would lower the secure key rate
dramatically. In our system, numerical simulation shows that the
secure key rate would drop from 0.3bit/pulse to 0.1bit/pulse.
\bibitem{CALIBRATION}
The channel efficiency $G$ was calibrated by using a strong laser
pulse and a power meter. To calibrate the total efficiency of Bob's
device $\eta$, a strong laser was feeded into Bob's system, while
the output of the photo diode was measured with a calibrated
transimpedance amplifier.
\bibitem{SEMICLASSICAL}
In this semiclassical picture, a coherent laser pulse can be treated
as a classical electromagnetic wave plus the vacuum noise. Since we
are estimating the excess noise (the noise above vacuum noise) here,
we can treat the leakage LE classically.
\bibitem{BB84_KEY}
In BB84 QKD with a perfect single photon source, the secure key rate
is given by
$R=\frac{1}{2}Q_{1}[1-f(e_{1})H_{2}(e_{1})-H_{2}(e_{1})]$. Here
$Q_1$ is the overall gain. $e_1$ is the QBER. $f(x)$ is the
bidirectional error correction efficiency and $H_{2}(x)$ is the
binary entropy function. Assuming $e_{1}=3\%$ and $f(e_{1})=1.22$, a
$10\%$ error on determining $e_{1}$ will result a $3\%$ change of
the secure key rate, which is tolerable. In decoy QKD, the equation
for calculating secure key rate is more complicated. Nevertheless,
as we showed in \cite{DECOY_THEORY2}, moderate errors on determining
system parameters are acceptable.
\bibitem{FIG4}
This is illustrated in Fig.4. Note the similarity between Fig.4(a)
(experimental results) and Fig.4(b) (simulation result under the
assumption of no excess noise).
\end{thebibliography}
\end{document}